\documentclass[12pt]{iopart}
\usepackage{cite}  
\usepackage[dvips]{graphics}
\usepackage{epsfig}

\begin{document}

\title[]{Generation of quasi static magnetic field in the relativistic laser-plasma interactions}

\author{Susumu Kato\dag
, Tatsufumi Nakamura\ddag, Kunioki Mima\ddag, Yasuhiko Sentoku\S, Hideo Nagatomo\ddag, and Yoshiro Owadano\dag}

\address{\dag\ National Institute of Advanced Industrial Science and Technology (AIST), Tsukuba, Ibaraki 305-8568, Japan}

\address{\ddag\ Institute of Laser Engineering (ILE), Osaka University, Suita, Osaka, 565-0871, Japan}

\address{\S\ Department of physics, University of Nevada, Reno, Reno, NV 89557, USA}

\begin{abstract}
The magnetic field generation by a relativistic laser light irradiated on a thin target at the oblique incidence is investigated using a two dimensional particle-in-cell simulation. 
The surface magnetic field inhibits the electron transport towards the inside plasma, when an incident angle exceeds the critical angle, which depends on the laser and plasma parameters.
\end{abstract}





\section{Introduction}
Laser plasma interaction in the relativistic regime, where the normalized vector potential $a_L\simeq (I_L\lambda_{\mu}^2/1.37\times10^{18})^{1/2}\geq1$, where $I_L$ and $\lambda_{\mu}$ are the laser intensity in $\textrm{W/cm}^2$ and the wavelength in $\mu$m, respectively, is crucial for inertial fusion energy with the fast ignitor scheme\cite{Tabak1994}, multi-MeV proton, and electron generation. In particular, the generation of magnetic fields during intense laser solid target interactions have attracted much interest for the past 30 years\cite{Haines1986,Wilks1992,Sentoku2000}. The magnetic fields are generated by various mechanisms, which are the currents produced from perpendicular density and temperature gradients in the ablated plasma\cite{Haines1986}, the radiation pressure associated with the laser pulse itself\cite{Wilks1992}, the current of fast electrons generated during the interaction, and the Weibel instability\cite{Sentoku2000}. Recently, self-generated magnetic fields, whose amplitudes are of the order of a few hundred mega gauss, have been observed in the overdense region of an irradiated solid target by a relativistic laser irradiation \cite{Tatarakis2002}. 

The self-generated magnetic fields can play a significant role in laser absorption \cite{Kruer1977} and transport of high energy particles in dense plasmas \cite{Sentoku2002}.
In addition, at the oblique incidence, the magnetic fields is localized on the plasma surface, which depth is less than a laser wavelength and amplitude is about one third of the laser electromagnetic field\cite{Brunel1988,Sentoku2004,nakamura2004,kato2004}. The magnetic field associated with a intense laser with intensity $I_L$ is extremely large, $B_L=290\times(I_L[\mathrm{W/cm}^2]/10^{19})$ MG, namely, the amplitudes are greater than 100 MG.
The generation of three kinds of quasistatic magnetic fields, which are located at the front surface, inner plasma, and rear surface in the interaction of ultrashort intense laser pulses with thin solid targets are reported in the previous paper\cite{kato2004}. First magnetic field which is located on a front surface arises very quickly, of which raising time is almost the same of the laser raising time. Second magnetic field is gradually growth in a inner plasma by the Weibel instability, which occurs between the fast electrons generated by the vacuum heating and their return currents. Final magnetic field that is located on the rear surface of the thin target arises also very quickly when electrons pass through the surface and return into the plasma. 

In the present paper, the magnetic field generation at the oblique incidence is investigated by a two dimensional particle-in-cell (2D PIC) simulation. We discuss the generation of the magnetic field and the electron transport inhibition due to the surface magnetic field. The surface magnetic field inhibits the electron transport towards a inside plasma, when an incident angle exceeds the critical angle, which depends on the laser and plasma parameters.

\section{2D PIC simulation}

We use the 2D PIC simulation with immobile ions. The schematic is shown in Fig 1. 
The plasma density $n_e=1.12\times10^{22}\textrm{cm}^{-3}$, which corresponds to $n_e/n_c=10$, where $n_c$ is the critical density for $\lambda=1\mu\textrm{m}$, and the initial temperature is $5$ keV.
The laser intensity rises in 5 fs and remains constant after that. The peak irradiated intensity $I=1\times10^{19}\textrm{W/cm}^2$, which corresponds to $a\simeq2.7$. The amplitude of magnetic field associated with a intense laser $B_L=290$ MG. In order to study the difference  in the magnetic fields of the incident angles, we simulate three incident angles $\theta=20^{\circ}, 60^{\circ}$ and $75^{\circ}$, respectively. The number of spatial grids and particles are $2048\times(416-1184)$ and $(4-11)\times10^6$, respectively. The simulation sizes depend on the incident angles.
In the simulations, all incident waves are p-polarized. The effects of the polarization have been studied in the previous paper \cite{kato2004}. The structures of the magnetic field of the s-polarized light are the same as that of the p-polarized light. The amplitude of the s-polarized light is less than 1/10 of the p-polarized light. 

The quasistatic magnetic fields $\widetilde{B}_z$ after 50 fs for incident angles $\theta=20^{\circ}$, $60^{\circ}$, and $75^{\circ}$ are shown in Figs. 2(a), 2(b), and 2(c), respectively. The quasistatic magnetic field is obtained by averaging over a laser period and the amplitude is normalized by that of the laser magnetic field. We find three kinds of quasistatic magnetic fields, which are located at the front and rear surfaces and inside plasma, for $\theta=20^{\circ}$ and $60^{\circ}$, as discussed in the previous paper. The normalized amplitude inner plasma is about one third of the laser electromagnetic field. This amplitude is well in agreement with Ref. \cite{Kruer2003}. The normalized amplitude at the front surface for $\theta=60^{\circ}$ is about half, namely, $\widetilde{B}_z\simeq150$MG. For $\theta=75^{\circ}$, three kinds of quasistatic magnetic fields are also observed. However, the amplitude of magnetic field inside the plasma is about 1/10 of the others. The reason of this small value is that the surface magnetic field inhibits the electron transport into the plasma and restricts the electron motion of the low energies at the surface. Since the restriction to the high energy electrons by the magnetic field is weak compared with the low energy ones, on the rear surface the magnetic field of the same kind as what was generated at an angle of others is generated by the high energy electrons.

\section{Concluding Remarks}

The magnetic field generation by the relativistic laser light irradiated on a thin target at the oblique incidence is investigated using a 2D PIC simulation code. We reconfirm the generation of the three kinds of quasistatic magnetic fields, which are located at the front and rear surfaces and inside plasma, for all incident angles of the simulation, as discussed in the previous paper \cite{kato2004}. However, the amplitudes are strongly depend on the incident angles. We find that the surface magnetic field inhibits the electron transport into the plasma and restricts the electron motion of the low energies at the surface for $\theta=75^{\circ}$.
When the incident angle is larger than a certain critical angle, a magnetic field will restrict a electron motion strongly.
The existence of the critical angle, which restricts the electron transport, is well consistent with the theory\cite{nakamura2004}.

\ack {This work was supported by Japan Society for the Promotion of Science (JSPS), KAKENHI(16540457).}

\newpage
\begin{figure}[htbp]
\begin{center}
  \includegraphics[width=150mm]{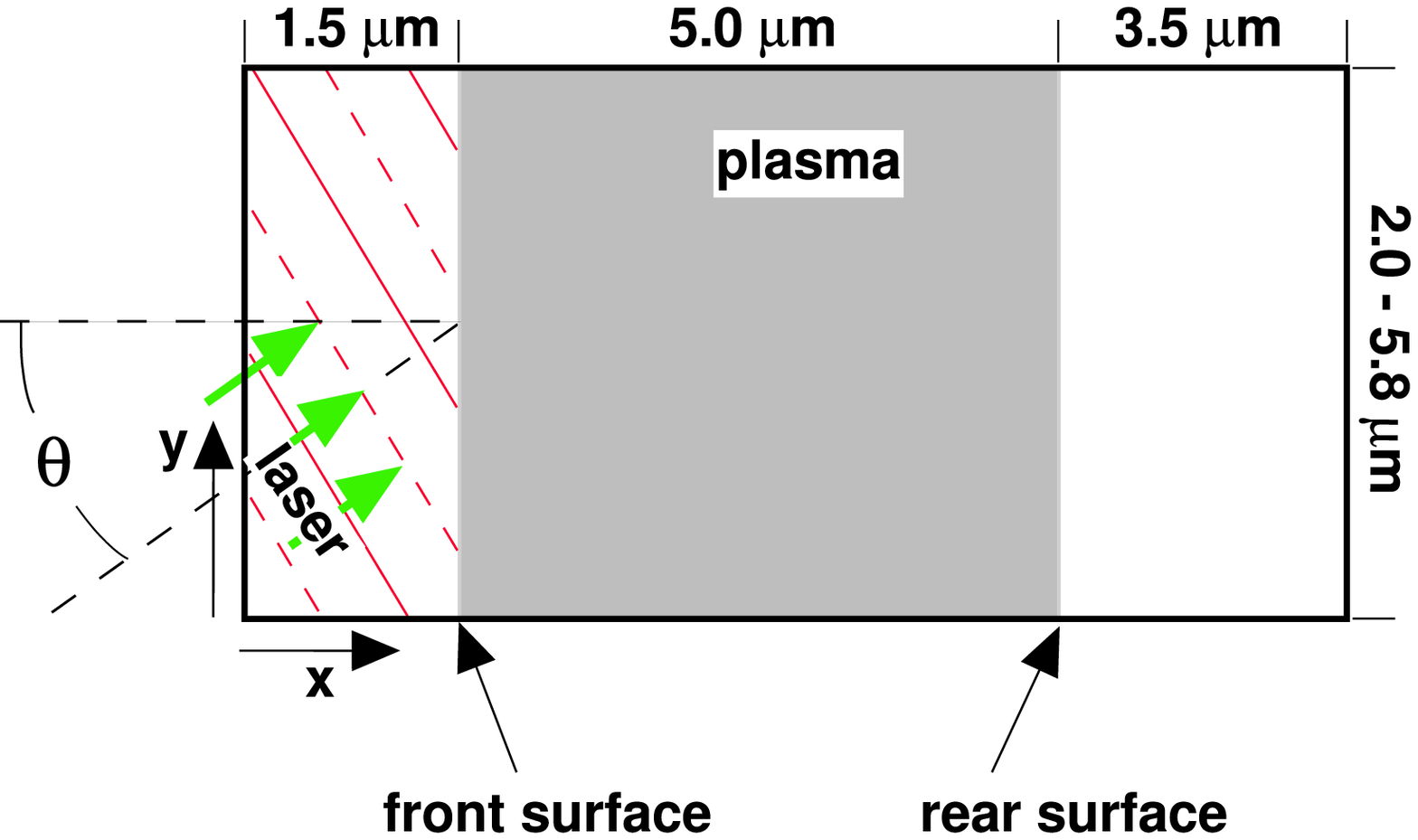}
\end{center}
\end{figure}
\Figure{\label{label1} A schematic of two dimensional PIC simulation. The density profile is initially homogeneous with a sharp plasma-vacuum interface. The incident laser light is periodic in y direction and two period in the simulation box. $\theta$ is an angle of incidence.}

\newpage
\begin{figure}[htbp]
\begin{center}
  \includegraphics[width=90mm]{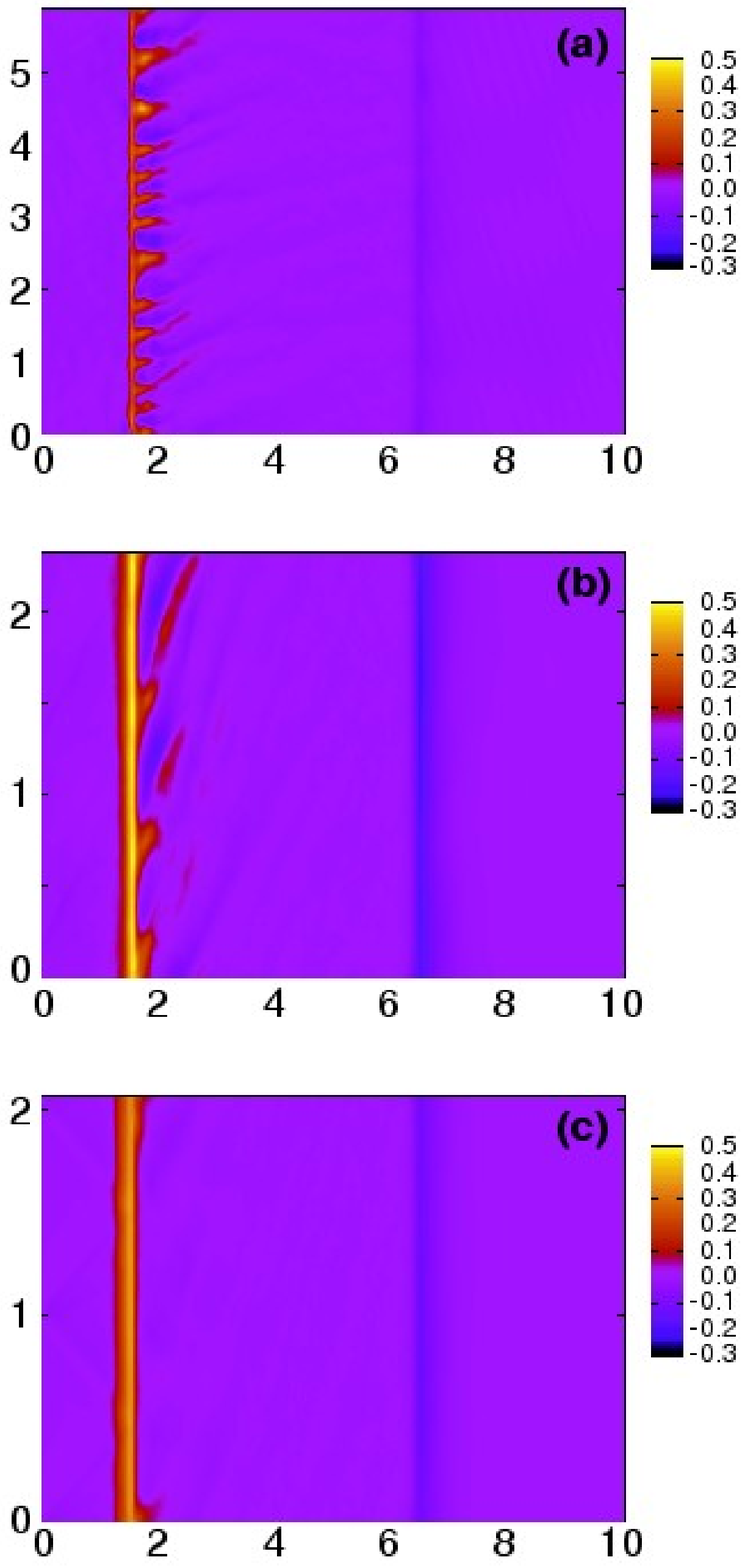}
\end{center}
\end{figure}
\Figure{\label{label2} Quasistatic magnetic fields $\widetilde{B}_z$ after $50$ fs for incident angles $\theta=20^{\circ}, 60^{\circ}$ and $75^{\circ}$, respectively. Spatial size is normalized by 1 $\mu$m.}


\section*{References}
\begin{thebibliography}{99}
\bibitem{Tabak1994}
M. Tabak {\it et al.}, Phys. Plasmas {\bf 1}, 1626 (1994);  R. Kodama {\it et al.}, Nature {\bf 412}, 798 (2001).
\bibitem{Haines1986}
M. G. Haines, Can. J. Phys. {\bf 64}, 912 (1986); J. A. Stamper, Laser Part. Beams. {\bf 9}, 841 (1991).
\bibitem{Wilks1992}
S. C. Wilks {\it et al.}, Phys. Rev. Lett. {\bf 69}, 1383 (1992); R. N. Sudan, Phys. Rev. Lett. {\bf 70}, 3075 (1993); R. J. Mason and M. Tabak, Phys. Rev. Lett. {\bf 80}, 524 (1998). 
\bibitem{Sentoku2000}
Y. Sentoku {\it et al.}, Phys. Plasmas {\bf 7}, 689 (2000). 
\bibitem{Tatarakis2002}
M. Tatarakis {\it et al.}, Nature {\bf 415}, 280 (2002); M. Tatarakis {\it et al.}, Phys. Plasmas {\bf 9}, 2244 (2002).
\bibitem{Kruer2003}
W. L. Kruer, Phys. Plasmas {\bf 10}, 2087 (2003). 
\bibitem{Kruer1977}
W. L. Kruer and K. Estabrook, Phys. Fluids {\bf 20}, 1688 (1977).
\bibitem{Sentoku2002}
Y. Sentoku {\it et al.}, Phys. Rev. E {\bf 65}, 46408 (2002); Y. Sentoku {\it et al.}, Phys. Rev. Lett. {\bf 90}, 155001 (2003). 
\bibitem{Brunel1988}
F. Brunel, Phys. Fluids {\bf 31}, 2714 (1988); H. Ruhl and P. Mulser, Phys. Lett. A {\bf 205}, 388 (1995).
\bibitem{Sentoku2004}
Y. Sentoku {\it et al.}, Phys. Plasmas {\bf 11}, 3083 (2004); K. Mima {\it et al.,} Proceedings of $19^{th}$ IAEA Fusion Energy Conference, Lyon, September, 2002.
\bibitem{nakamura2004}
T. Nakamura {\it et al.}, Proceedings of the Third International Conference on Inertial Fusion Sciences and Applications (IFSA2003), Editors:  B. A. Hammel  D. D. Meyerhofer  J. Meyer-ter-Vehn  H. Azechi, p.478 (American Nuclear Society, Inc., 2004).
\bibitem{kato2004}
S. Kato {\it et al.}, Journal of Plasma and Fusion Research SERIES Vol.6  (to be published).
\bibitem{Kruer1985}
W. L. Kruer and K. Estabrook, Phys. Fluids {\bf 28}, 430 (1985).
\bibitem{Brunel1987}
F. Brunel, Phys. Rev. Lett. {\bf 59}, 52 (1987).

\end{thebibliography}
\end{document}